\documentclass[aps,preprint,prl,superscriptaddress,showpacs,amsmath,amssymb]{revtex4}% Physical Review

\usepackage{graphicx,color}% Include figure files
\usepackage{dcolumn}% Align table columns on decimal point
\newcolumntype{.}{D{.}{.}{-1}}
\usepackage{bm}% bold math
\usepackage{epstopdf}
\usepackage{subfigure}

\newcommand{\lmp}{LiMnPO\(_4\)}
\newcommand{\lfp}{LiFePO\(_4\)}

\newcommand{\lmfp}{Li(Mn,Fe)PO\(_4\)}
\newcommand{\lmfpx}{LiMn\(_{1-x}\)Fe\(_x\)PO\(_4\)}
\newcommand{\lmnp}{LiMn\(_{1-x}\)Ni\(_x\)PO\(_4\)}
\newcommand{\lico}{LiCoPO\(_4\)}

\newcommand{\tn}{$T_{\rm N}$}
\newcommand{\tm}{$T_{\rm m}$}
\newcommand{\tz}{$T_{\rm z}$}
\newcommand{\vz}{$V_{\rm z}$}
\newcommand{\etal}{\textit{et~al.}}

\begin{document}

%\begin{frontmatter}

\title{High-pressure optical floating-zone growth of Li(Mn,Fe)PO$_4$ single crystals}

\author{Christoph Neef*}
\affiliation{Kirchhoff Institute of Physics, Heidelberg University, D-69120 Heidelberg, Germany}
%\cortext[corrauth]{Corresponding author}
\email{christoph.neef@kip.uni-heidelberg.de}
\author{Hubert Wadepohl}
\affiliation{Anorganisch-Chemisches Institut, Heidelberg University, D-69120 Heidelberg, Germany}
\author{Hans-Peter Meyer}
\affiliation{Institut f\"{u}r Geowissenschaften, Heidelberg University, D-69120 Heidelberg, Germany}
\author{R\"{u}diger Klingeler}
\affiliation{Kirchhoff Institute of Physics, Heidelberg University, D-69120 Heidelberg, Germany}
\affiliation{Centre for Advanced Materials, Heidelberg University, D-69120 Heidelberg, Germany}

\begin{abstract}
Mm-sized \lmfpx\ single crystals with $0\leq x \leq 1$ were grown by means of the traveling floating-zone technique at elevated Argon pressure of 30~bar. For the various doping levels, the growth process was optimized with respect to the composition-dependant effective light absorption and transparency of the materials. A convex crystal/melt interface, determined by the angle of incident light, was identified to be particularly crucial for a successful growth. The resulting large single crystalline grains are stoichiometric. Structure refinement shows that lattice parameters as well as the atomic positions and bond lengths linearly depend on the Mn:Fe-ratio. Oriented cuboidal samples with several mm$^3$ of volume were used for magnetic studies which imply an antiferromagnetic ground state for all compositions. The N\'eel-temperature changes from \tn\ = 32.5(5)\,K in \lmp\ to 49.5(5)\,K in \lfp\ while the easy magnetic axis in the ordered phase flips from the crystallographic $a$- to the $b$-axis upon Fe-doping of $x<0.2$.
\end{abstract}

\maketitle

%\begin{keyword}
%A2.Single crystal growth \sep A2.Floating zone technique \sep B1.Phosphates \sep B2.Lithium-ion battery materials \sep B2.Magnetic materials
%\end{keyword}

%\end{frontmatter}

%% \linenumbers

\section{Introduction}

Olivine-structured orthophosphates Li\textit{M}PO$_4$ \textit{M} = (Mn, Fe, Co, Ni)~\cite{Padhi1997,Yamada2006,Neef2013,Neef2015} are in the focus of both fundamental and applied research as they offer various desired functionalities. One prominent property is their strong potential for electrochemical energy storage applications in lithium-ion secondary batteries. While the technological relevance of this field stimulates further research efforts to obtain a deeper understanding of the principles of lithium exchange in Li\textit{M}PO$_4$, there is only a very limited number of compounds for which single crystals were synthesized. This strongly hinders detailed investigations of the materials, among which anisotropic electric and Li-transport properties are probably the most relevant ones for understanding, modelling, and optimizing the materials for lithium-ion batteries.

Another sought class of advanced functional materials shows strong coupling of electric and magnetic degrees of freedom which in the Li-based olivine orthophosphates appears, e.g., in large magnetoelectric effects~\cite{Vaknin2004,Toft-Petersen2012}. In the particular case of multiferroics, this coupling drives the coexistence of electric, structural and/or magnetic orders in the same phase. Hence, in multiferroic materials, one may control the magnetic response by electric fields or stress, or use magnetic fields to tune the electric and structural properties. Olivine phosphates provide a new route in this field of multiferroics as \lico\ shows a new form of ferroic order, namely ferrotoroidicity, which adds to widely known ferromagnetism, ferroelectricity, and ferroelasticity~\cite{vanAken2007}. In the ferroic state, small domains of uniformly aligned electric, magnetic, and toroidal moments are formed which microscopically change upon switching the ferroic state. Therefore, proper understanding of the formation of such multi-domain structures is needed to fully exploit ferroic orders, and single-crystalline model systems are mandatory to elucidate this phenomenon.

The work at hand addresses the doping series \lmfpx , including the end members \lfp\ and \lmp . \lfp\ is an commercially well-established cathode material with good thermal stability and decent power density. Due to the higher electrochemical potentials, replacing Fe by Mn, Co, or Ni promises increased energy densities. Application of the high-voltage compounds LiCoPO$_4$ and LiNiPO$_4$ is, however, still hindered by degradation issues and a lack of stable electrolytes. We have hence focussed at Fe-substitution in \lfp\ by Mn in order to investigate the doping series \lmfpx . In these materials, the established battery cell designs yield stable cycling behavior while the cell voltage can increase up to 4.1\,V vs. Li/Li$^+$ in \lmp . Furthermore, the \lmp\ structure is reported to be less susceptible towards Li/\textit{M} antisite disorder than \lfp , protecting the 1D Li-diffusion channels which are supposed to be relevant for the Li-exchange from blockage~\cite{Jensen2013}. A feasible application of \lmp\ seems however difficult so far, since ionic and electronic conductivities are rather low due to Jahn-Teller distortions in the delithiated phase MnPO$_4$~\cite{Seo2010}, and the electron polaron conduction mechanism suffering from high energy barriers~\cite{Ong2011,Johannes2012}. Only partial replacement of Fe by Mn in \lmfp\ is hence a promising approach to concomitantly achieve higher cell voltages and high power capability as was shown by several studies, so far however limited to polycrystalline samples~\cite{Molenda2013,Koentje2014,Martha2009}.

In order to investigate in detail intrinsic structural, thermodynamic, and transport properties, large and high-quality single crystals are necessary in particular to elucidating any effects showing anisotropy. In addition, single crystals are needed to investigate structural properties and defects, which can best be seen in single crystal XRD experiments, and of their interplay with ionic conduction. Recently, there have been a few reports on the growth of \lmp\ and \lfp\ crystals either by flux methods~\cite{Li2006,Janssen2013} or by the optical floating-zone technique~\cite{Chen2005,Wizent2009,Wizent2011,Wang2013}. Mixed transition-metal single crystals have not been reported yet. Even the existing reports on the end members however yield contradicting results on the ionic conductivity which implies that the crystal properties are significantly affected by the actual synthesis approach and growth conditions. Availability of high quality samples, reflecting the intrinsic bulk properties, is therefore crucial for any studies of fundamental relations.

The present work reports on the optical floating-zone growth of \lmfpx\ single crystals with $x$~=~0, 0.1, 0.2, 0.3, 0.5 and 1. The growth was performed upon application of a pressure of 30~bar of purified Ar-atmosphere. Elevated-pressure synthesis was motivated by previous studies which have shown that application of pressure minimizes Li$_2$O- and Mn-evaporation from the melt~\cite{Yang2010,Prabhakaran2002}. In addition, usage of purified Ar implies the absence of residual oxygen in the growth chamber and hence prevents oxidation of Fe$^{2+}$ to Fe$^{3+}$. For the various doping levels, the growth process had to be optimized with respect to changes of the relevant material parameters, such as the transparency of the crystal and the melt or the melting temperature. Thereby, the shape of the interface between the crystal and the melt was identified to be of crucial importance for proper grain selection and crystal growth. In order to ensure a convex interface, specific effort was put into adjusting the geometry of the vertical growth-furnace used. A method of regulating the angle of incidence and distribution of light and thereby the interface shape is discussed with respect to the resulting grain-selection properties and crystal qualities. The resulting mm-sized single crystalline grains are stoichiometric and our structural investigations of the doping series show a solid-solution behavior between the end members. In addition, there is moderate tendency to Li-\textit{M} antisite disorder which seems to be more pronounced for \lfp\ as compared to \lmp . Magnetic characterization reveals highly anisotropic antiferromagnetic ordering and variation of \tn\ upon doping. Sharp $\lambda$-like anomalies of the magnetic specific heat are observed which confirm the high-quality of the single crystals.

\section{Experimental}

Polycrystalline starting materials \lmfp\ were prepared by a solid-state reaction route using Lithium-carbonate, Manganese-carbonate, Iron-oxalate-dihydrate and Ammonium-dihydrogen-phosphate. All precursors were mixed in the desired ratio, while the stoichiometry of Li:\textit{M}:PO$_4$ (\textit{M}= Mn,Fe) was always kept at 1:1:1. The starting mixtures were wet-ground in acetone in a ball mill for 3~h using agate media. Consecutively, the product was dried under low pressure argon atmosphere at 60\,°C for some hours, decarbonated at 370\,°C for 12\,h and re-ground. The final sintering step was carried out at temperatures ranging between 650\,°C ($x=1$) and 750\,°C ($x=0$). Here, the temperature was increased with a rate of 300\,°C/h at a pressure of 10\,mbar and under Ar-flow of 250\,sccm in order to remove any gaseous reaction products. After reaching the target temperature, pressure was increased to atmospheric pressure (1000\,mbar) in order to minimize Li-evaporation. The resulting reaction products were re-ground, pelletized and heated again at the same temperatures for 12\,h at a pressure of 1500\,mbar without continuous Ar-flow. The final products exhibit white to light gray colour. Finally, feed rods of 7\,mm diameter were fabricated and isostatically pressed at 2\,kbar.

The crystal growth was carried out in a high-pressure floating zone furnace (HKZ, SciDre)~\cite{NeefDiss}. For the work reported, a 5\,kW xenon arc lamp was used which light is focussed on the sample by means of two confocal mirrors in vertical arrangement~\cite{Souptel2007,Behr2007}. The light intensity is adjusted by blocking or opening the optical path-way by means of a four leave mechanical iris. A quartz growth chamber of 72~mm in length and with a wall thickness of 14\,mm was used. Argon 5.0 pressure of 30\,bar was applied at a flow-rate of 0.125\,l/min. $In-situ$ temperature profiles of the sample and melting zone were obtained stroboscopically by means of a two-colour pyrometer which can me moved vertically along the growth chamber during the growth process. Feed- and seedrod were counter rotated at 27 and 19\,rpm, respectively, in order to ensure mixing of the melt. Pulling rates of 3 to 5~mm/h were applied during all growth experiments. Further characteristics of the growth process and the effect of the growth parameters will be discussed in section~\ref{crystalgrowth}.

EDX analysis of all resulting crystals was done by means of an Oxford Leo 440 scanning electron microscope equipped with an Inca X-Max 80 detector. The acceleration voltage was 20\,kV, the working distance was 25\,mm, and the counting time was 100\,s (lifetime) at about 10,000\,cps. Further experimental information can be found in the supporting information. Both the polycrystalline starting materials and the ground single crystals were studied by powder X-ray diffraction measurements on a Bruker D8 Advance ECO diffractometer equipped with a Cu-anode and an SSD-160 line-detector in Bragg-Brentano geometry. The measurement range was 10° to 70° with a step-width of 0.02° and an integration time of 32\,s per step. Rietveld refinement of the patterns and calculation of the lattice parameters were done using the FullProf Suite 2.0~\cite{fullprof}. Single-crystal X-ray investigations were carried out either using a Bruker AXS Smart 1000 CCD diffractometer (sealed X-ray tube, graphite monochromator), Rigaku Supernova Dualflex-AS2 CCD diffractometer (microfocus X-ray tube, multilayer mirror optics) or an Agilent Technologies Supernova-E CCD diffractometer (microfocus X-ray tube, multilayer mirror optics) at a stabilized low temperature. For detailed experimental description, see the supporting information. X-Ray Laue diffraction in back scattering geometry was used to orient the single crystals which were then cut to cuboids with respect to the crystallographic main directions using a diamond-wire saw. The static magnetic susceptibility of the single crystalline cuboids was investigated by means of SQUID magnetometry (Quantum Design MPMS-XL5) at a field strength of $\mu _0 H$~=~0.1\,T.

\section{Crystal growth}\label{crystalgrowth}

We report the successful growth of high-quality \lmfpx\ single crystals with $0\leq x \leq 1$. In the following, characteristics of the growth procedure are discussed with a focus on optimization of the growth process. Due to different properties of the materials with different doping levels $x$, the growth parameters had to be optimized accordingly. As will be discussed in the following, the main impact arises from the different melting temperatures \tm\ of the materials and from the strong differences in light absorption. Due to the insufficient grain selection, crystalline grains are relatively small (about 10\,mm$^3$) in the \lmp -rods and the resulting material is relatively brittle. For \lfp , the grain size about 0.5\,cm$^3$) is restricted by a drain of melt and strong horizontal grain growth. The composition of LiMn$_{0.7}$Fe$_{0.3}$PO$_4$ turned out to be the optimum composition in terms of grain size and growth stability, combining good light absorption in the zone and a slightly convex interface with the absence of feed rod fracturing. In this way, crack free single crystalline parts with a length of several cm (volume of about 1\,cm$^3$) were obtained.
The grown rods are of orange colour in the case of \lmp\ which changes to yellow for Fe-contents between 10 and 30\,\%, brown for 50\,\%, and eventually dark green for \lfp\ (see the supporting information).

\subsection{Melting temperatures and temperature profile along the melting zone}

\begin{figure}[h]
\includegraphics [width=0.7\columnwidth,clip] {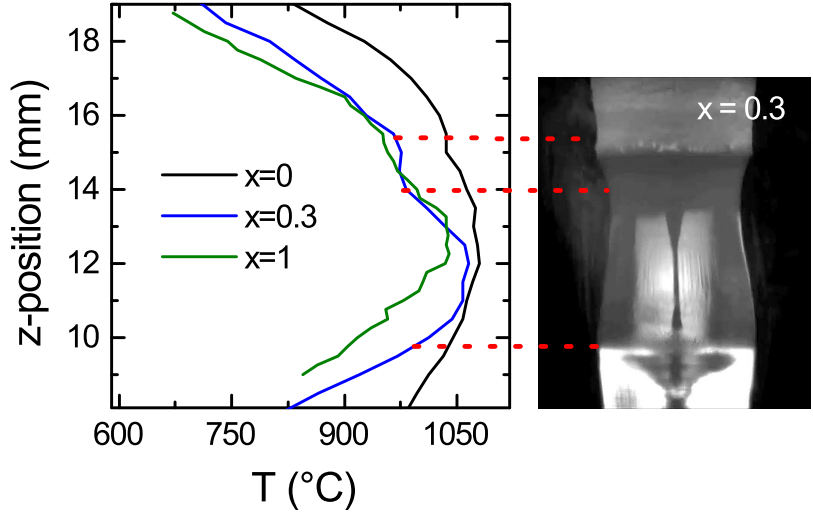}
\caption{Vertical temperature profiles showing a zonal plateau and a melt/solid-coexistence plateau, respectively a shoulder, near the feed rod. The image of a LiMn$_{0.7}$Fe$_{0.3}$PO$_{4}$ sample during the growth illustrates the different sections (see the text).} \label{TempProf}
\end{figure}

\begin{figure}[!h]
\includegraphics [width=0.7\columnwidth,clip] {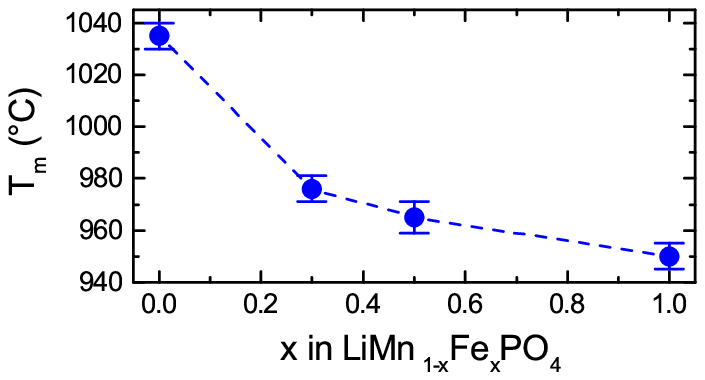}
\caption{Melting temperatures \tm\ $vs.$ doping level $x$ in \lmfpx .} \label{temp-comp}
\end{figure}

During all growth experiments, the heating power and the temperature of the melt were kept as small as possible in order to avoid material evaporation and decomposition~\cite{Chen2005}, but still high enough to ensure stable growth. Accordingly, the zone temperatures were adjusted to values between 1080~°C for \lmp\ and 1040°C for \lfp . In order to illustrate the similarities and the differences for the different doping levels, the temperature profiles at and around the melting zones are shown for the examples of \lmfpx\ with $x$~=~0, 0.3, 1 (Fig.~\ref{TempProf}). All profiles exhibit a bell-like shape. The melting zones are characterized by a broad and only slightly curved central region and a small plateau-like shoulder. The broad central region resembles the temperature of the melt. A small shoulder slightly above displays the interface between the feed rod and the melting zone which in the picture of the LiMn$_{0.7}$Fe$_{0.3}$PO$_{4}$ sample in Fig.~\ref{TempProf} shows up as a grey region above the position of the minimal zone radius.
In this region, there is a coexistence of melt and solid so that it reflects the actual melting temperature \tm\ of the material. For the different compositions under study, we obtain $T_{\rm m} = 1035\pm 5^{\circ}$C ($x=0$), $976\pm 5^{\circ}$C ($x=0.3$), $965\pm 6^{\circ}$C ($x=0.5$), and $950\pm 5^{\circ}$C ($x=1$) by reading off the temperatures at the shoulder (see Fig.~\ref{temp-comp}). For $x=0$ and $x=1$, the results agree to the melting temperatures reported in Ref.~\cite{Wizent2011} and \cite{Chen2005}. The temperature profiles in Fig.~\ref{TempProf} also enable determining the temperature gradients which are, e.g., about 40\,°C/mm in \lmp\ and 70\,°C/mm in \lfp .

%%%%%%%%%%%%%%%%%%%%%%%%%%%%%%%%%%%%%%FIG%%%%%%%%%%%%%%%%%%%%%%%%%%%%%%%%%%%%%%%%%%%%%%%%%%%%%
%%%%%%%%%%%%%%%%%%%%%%%%%%%%%%%%%%%%%%%%%%%%%%%%%%%%%%%%%%%%%%%%%%%%%%%%%%%%%%%%%%%%%%%%%%%%%%

%%%%%%%%%%%%%%%%%%%%%%%%%%%%%%%%%%%%%%FIG%%%%%%%%%%%%%%%%%%%%%%%%%%%%%%%%%%%%%%%%%%%%%%%%%%%%%
%%%%%%%%%%%%%%%%%%%%%%%%%%%%%%%%%%%%%%%%%%%%%%%%%%%%%%%%%%%%%%%%%%%%%%%%%%%%%%%%%%%%%%%%%%%%%%

\subsection{Zone shape and growth behaviour}

\begin{figure*}[h]
\centering
\includegraphics [width=0.95\columnwidth,clip] {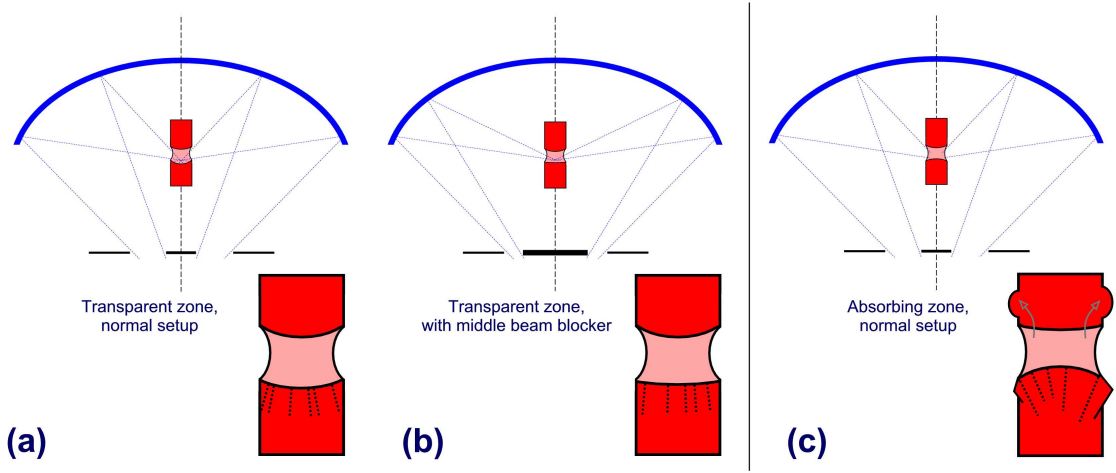}
\caption{Schematic sketch of the upper part of the floating zone furnace (HKZ). The light pathways and crystal/melt interfaces are shown for several optical arrangements and zonal light absorbabilities along with the resulting grain growth. (a) For a transparent melting zone, the vertical light flux results in a concave interface. (b) Blocking the vertical, near-center light flux parallel to the optical axis yields flattening of the interface curvature. (c) A convex interface is formed in the case of a highly absorbing zone.} \label{hkz-geometry}
\end{figure*}

Due to the different specific light absorption appearing in dark melting zones in the Fe-rich compounds and almost transparent zones in the Mn-rich ones, the optical setup of the image furnace had to be separately adjusted for each composition. The general behaviour is illustrated by Fig.~\ref{hkz-geometry} which shows the vertical setup (upper mirror and optical axis) and the corresponding light flux.
The sketches schematically illustrate how the shape of the melting zone is influenced by both, the absorbability of the zone and the geometry of the optical setup. Fig.~\ref{hkz-geometry} (a) shows the case of a highly transparent zone as it is realised in \lmp . Due to the weak absorption in the melt, light passes the zone and concave crystal/melt and melt/feed rod interfaces are formed, caused by the non-horizontal light flux and direct heating of the rods.
Such a concave interface on the crystal, however, is unfavorable for proper grain selection and hence the growth of large single crystalline grains is hindered. In order to enable proper grain selection for the Mn-rich compounds, the setup was hence changed as shown in Fig.~\ref{hkz-geometry} (b). A so-called 'beam-blocker' was placed in the center of the optical path in order to particularly screen the vertical near-center light flux parallel to the optical axis from the sample. This measure reduces the heating power on the crystal/melt interface of the seed-rod and yields less concave interfaces which facilitate the growth of large single crystalline grains.

The situation is different for a highly absorbing zone as it is realised in \lfp . Here, light is rather completely absorbed inside the zone and does not heat the seed-rod directly (see Fig.~\ref{hkz-geometry} (c)). Thus, the shape of the interface can develop undisturbed which leads to a convex crystal/melt interface. The convex shape promotes horizontal growth and grain selection, which permits large single crystalline volumes. Note, that in the doping series \lmfpx\ the concentration $x$~=~0.3 can already be considered as Fe-rich in this respect, as its melting zone is dark and highly absorbing.

%%%%%%%%%%%%%%%%%%%%%%%%%%%%%%%%%%%%%%FIG%%%%%%%%%%%%%%%%%%%%%%%%%%%%%%%%%%%%%%%%%%%%%%%%%%%%%
%%%%%%%%%%%%%%%%%%%%%%%%%%%%%%%%%%%%%%%%%%%%%%%%%%%%%%%%%%%%%%%%%%%%%%%%%%%%%%%%%%%%%%%%%%%%%%

\begin{figure*}[!h]
\centering
\includegraphics [width=0.95\columnwidth,clip] {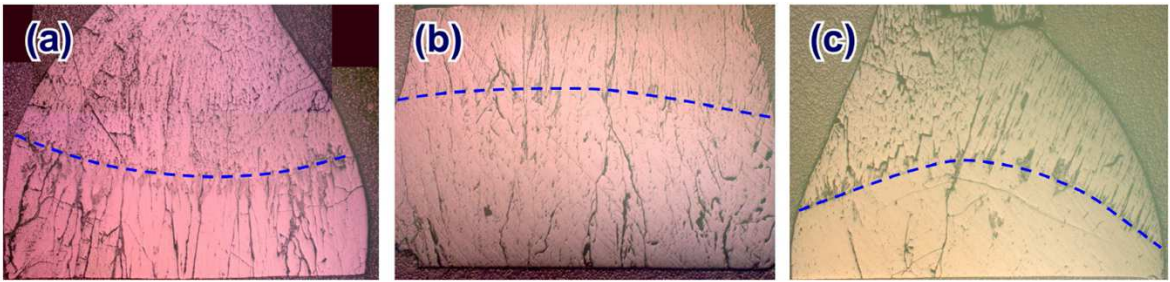}
\caption{Microscopy images of the frozen zones of \lmp\ (a), LiMn$_{0.7}$Fe$_{0.3}$PO$_{4}$ (b) and \lfp\ (c). The curvature of the crystal/melt interface changes in accordance with the different absorbabilities of the zones (dashed lines).} \label{zones}
\end{figure*}

Direct experimental information on the interface shape can be obtained from the frozen melting-zones at the end of the growth regions. Fig.~\ref{zones} shows microscopy images of the zones for $x=0$ (a), $x=0.3$ (b), and $x=1$ (c) after a growth process without using a middle-beam blocker. The images show how the shape of the interface changes from concave to convex upon increase of the Fe-concentration as it is expected from the increasing absorption of the melt in the Fe-rich compounds. The different absorbabilities of the melts are also reflected in the temperature profiles in Fig.~\ref{TempProf}. In \lfp , there is zone-centered heating which results in a smaller zone and a somehow sharper temperature profile. Consequently, the associated temperature gradient is rather high. In the case of Mn-rich materials with nearly transparent melting zones, the temperature profile is much flatter as heating is more rod accentuated.

%%%%%%%%%%%%%%%%%%%%%%%%%%%%%%%%%%%%%%FIG%%%%%%%%%%%%%%%%%%%%%%%%%%%%%%%%%%%%%%%%%%%%%%%%%%%%%
%%%%%%%%%%%%%%%%%%%%%%%%%%%%%%%%%%%%%%%%%%%%%%%%%%%%%%%%%%%%%%%%%%%%%%%%%%%%%%%%%%%%%%%%%%%%%%

Hence, for the growth of materials with $x=0, 0.1$ and $0.2$, a circular shaped middle-beam-blocker was introduced into the light-pathway near the focus point of the upper mirror. This leads to a shielding of the fraction of light with highest-incidence angles from the upper mirror and thus from the sample. The resulting loss of total light intensity and heating power, which we roughly estimate to be around 30\,\%, was compensated for by increasing the primary power of the lamp. Applying this method, single crystalline grains of several mm$^3$ of volume were successfully obtained for the Mn-rich compounds.

While concave interfaces prohibit grain selection, a strongly convex interface leads to the growth of cone-shaped grains in \lfp\ (see Fig.~\ref{hkz-geometry} (c)). Such a copped interface shape may yield fracturing of the seed-rod as at low growth speed of few mm/h external grains will not diminish due to expanding internal grains, but may leave the generally circular feed rod shape by growing in horizontal direction. Such behaviour is feasible due to the good wetting properties of the Fe-rich melts, but it can cause unfavorable changes of the zone shape and volume.
An additional consequence of the strong wetting behaviour is seen on the feed rod as the capillary effect of the pressed powder causes an upward flux of melt out of the zone. Our experiments show re-solidification of the melt outside the illuminated and heated area which results in the aggregation of material on the outside of the feed rod. This phenomenon leads to fluctuations of the zone volume, since the upward flux initially decreases the volume of the zone while the subsequent re-melting of the aggregated material enlarges the melting zone again.

Therefore, stable growth was achieved only by pre-melting the initial feed rod with a seed-rod pulling-rate higher than the feed rod feeding-rate. This removes remaining cavities in the pressed powder, thereby diminishing capillary effects due to the higher density of pre-molten material. In addition, the reduction of rod-diameter from initially 7\,mm (powder) down to about 4.5\,mm (pre-molten, fastly pulled) allows a higher feeding-rate in the final growth process, leading to less variation in the zone volume because the feed rods are molten faster so that upward flux of the melt is suppressed.

\subsection{Zone stability}

In addition to the interface shape and grain selection discussed above, the interplay of melt transparency and zone-volume has a big influence on the growth stability, too, and particularly affects the self-adjusting properties of the melting-zone. This will be discussed by considering the response of the growth to a change of zone-volume \vz\ at constant light flux. Such a destabilizing event might occur due to an inhomogeneous feed-rod density and thus a nonuniform feed of material into the melt or strong horizontal crystal growth, as was observed for \lfp , which might cause an increased drain of melt from the zone. The associated effects on the cooling and heating power $P_{\rm cool}$ and $P_{\rm heat}$, respectively, in the melting zone are discussed below.

Irrespective of the composition, the increase of \vz\ will lead to stronger cooling of the zone because heat transfer by radiation and convection increases with the surface area of the zone, yielding $P_{\rm cool} \propto V_{\rm z}^{1/2}$. In contrast, as discussed above, light absorption strongly changes with $x$. In the Fe-melts, the total absorbed heating power is rather independent on \vz\ as incident light is completely absorbed even in small melting zones, i.e. $P_{\rm heat}^{\rm Fe} \propto \rm{const}$. As a consequence, the melting zone will become colder upon increasing \vz .
On the other hand, a decrease in the zone temperature \tz\ will shift the solid-melt boundaries towards the zone-center as the temperature at the boundary of the melt \tz\ becomes smaller than \tm\ thereby decreasing \vz\ again. The process is hence self-stabilizing and robust against minor disturbances. Although, a heavy drain of melt due to seed-rod fracturing or feed rod re-solidification as discussed above cannot be compensated for. In our experiments, this effect in several cases interrupted the growth process after a few cm of crystal growth.

In contrast, the transparent zones in Mn-rich melts do not absorb all of the incident light. Increasing \vz\ will hence enhance the total light absorption and thus the heating power according to the Beer-Lambert law, i.e. \newline $P_{\rm heat}^{\rm Mn} \propto (1-{\rm{exp}}[-\epsilon V_{\rm z}^{1/2}])$, with $\epsilon$ being a constant. In particular, the increase of $P_{\rm heat}^{\rm Mn}$ exceeds the additional cooling so that \tz\ will increase upon enlarging the zone volume. In contrast to the Fe-rich melts, the solid/melt-boundaries accordingly shifts towards the feed and seed rod, respectively, thereby increasing the zone-volume even further. The process is self-amplifying. Thermal runaway is however prevented by the geometric confinement of the incident light. The maximum zone length and zone volume are therefore restricted by the light distribution. In our experiments, sufficient stability upon growing Mn-rich materials was achieved by using relatively high zone temperatures. In this case, the solid-melt boundary is rather determined by the incident light distribution than by the zone temperature \tz .

\section{Crystal quality and crystallographic properties}

%%%%%%%%%%%%%%%%%%%%%%%%%%%%%%%%%%%%%%FIG%%%%%%%%%%%%%%%%%%%%%%%%%%%%%%%%%%%%%%%%%%%%%%%%%%%%%
\begin{figure}[!h]
\centering
\includegraphics [width=0.7\columnwidth,clip] {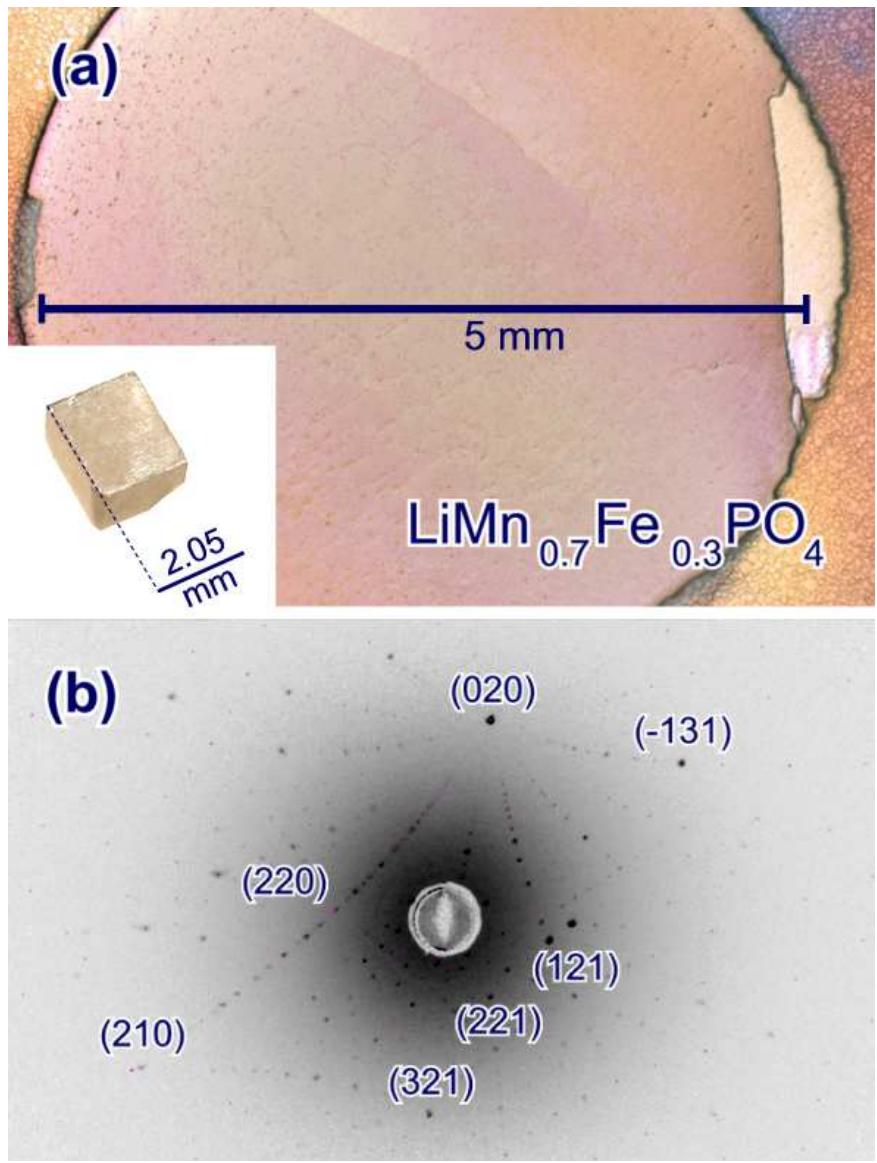}
\caption{(a) Light microscopy image of an as-grown LiMn$_{0.7}$Fe$_{0.3}$PO$_{4}$ rod, cut perpendicular to the growth direction, and (b) the corresponding Laue pattern (black spots) and simulation (purple spots). The growth direction is tilted by approximately 25° from the crystallographic $b$-axis. The inset in (a) shows an oriented cuboidal single crystal of $2.5 \times 2.05 \times 1.95$\,mm$^3$.} \label{GrowthDirection}
\end{figure}
%%%%%%%%%%%%%%%%%%%%%%%%%%%%%%%%%%%%%%%%%%%%%%%%%%%%%%%%%%%%%%%%%%%%%%%%%%%%%%%%%%%%%%%%%%%%%%

Several samples were cut from the grown rods and analyzed by optical microscopy and by Laue XRD. Fig.~\ref{GrowthDirection} (a) shows a  light microscopy image of a cut perpendicular to the growth direction, taken from a LiMn$_{0.7}$Fe$_{0.3}$PO$_4$ sample, along with the corresponding Laue image (Fig.~\ref{GrowthDirection} (b)). The cross section shows several large grains. The Laue pattern of the large grain, which covers the center of the section, confirms high crystallinity of the sample. It can be indexed by an orthorombic lattice with $Pnma$-symmetry which is characteristic for olivine-like phosphates~\cite{Yamada2006}. The corresponding simulated pattern is shown in the figure by overlayed purple spots. The growth direction is tilted by approximately 25° from the crystallographic $b$-axis. A similar behaviour is found for \lmp\ and \lfp , implying that a preferred growth direction near [010] is universal for the whole series. The grown rods were used to prepare cuboidal samples with several mm$^3$ in volume and faces cut perpendicular to the crystallographic main directions (one example is shown in the inset of Fig.~\ref{GrowthDirection} (a)). Single crystallinity was asserted by confirming the equallity of Laue-patterns from opposing sample faces (see supporting information).

The elemental composition for each specimen was determined by energy-dispersive X-ray spectrometry in ten single-point measurements on a polished surface. The results confirm the element ratio to be close to the composition of the starting materials (see the supporting information. Note, that the content of Li can not be obtained by this technique. LiO$_{0.5}$/f.u. was assumed for this analysis). The measured Fe:Mn ratios are summarized in Table~\ref{tabstoich}. The deviations of the ten point-measurements from their mean value are less than 0.5\,\% which suggests a very homogeneous distribution throughout the sample volumes. There are no visible systematics in the deviation of the nominal and the measured compositions.

\begin{figure}[!h]
\centering
\includegraphics [width=0.7\columnwidth,clip] {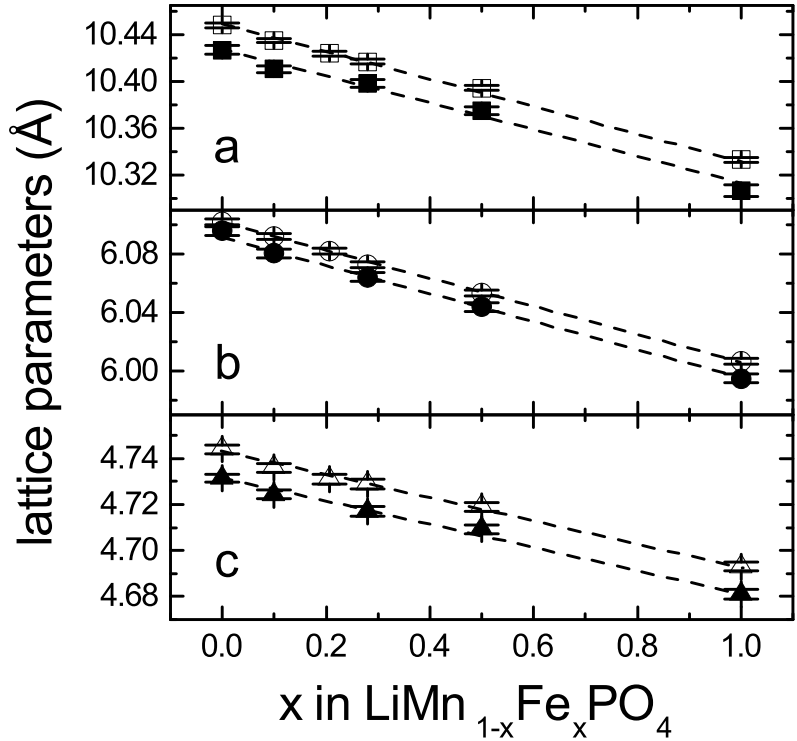}
\caption{Evolution of the lattice parameters in \lmfpx\ upon variation of the Fe-content $x$. Open symbols: powder XRD data (taken at 300\,K), filled symbols: single crystal XRD data (taken at 100\,K). Dashed lines are guides to the eyes.} \label{latticeConst}
\end{figure}

Phase purity of the samples was confirmed by powder XRD studies taken at room temperature from reground parts of the crystalline rods. All patterns can be indexed in the $Pnma$-symmetry. The data do not show any traces of impurity phases. In addition, Rietveld refinement (see the supporting information) of the powder XRD patterns enables evaluating the composition dependence of the lattice parameters (Fig.~\ref{latticeConst}). For all three main crystallographic directions, the lattice constants $a$, $b$, and $c$ show linear decrease upon changing the doping level $x$, demonstrating the solid solution behaviour between \lmp\ and \lfp . The shift of lattice constants agrees well with the difference of high-spin ionic radii in six-fold coordination for Mn$^{2+}$ (0.83\,\AA ) and Fe$^{2+}$ (0.78\,\AA )~\cite{Shannon1976}.

%%%%%%%%%%%%%%%%%%%%%%%%%%%%%%%%%%%%%%FIG%%%%%%%%%%%%%%%%%%%%%%%%%%%%%%%%%%%%%%%%%%%%%%%%%%%%%
%%%%%%%%%%%%%%%%%%%%%%%%%%%%%%%%%%%%%%%%%%%%%%%%%%%%%%%%%%%%%%%%%%%%%%%%%%%%%%%%%%%%%%%%%%%%%%

\section{Single-crystal structure refinement}

Single crystal X-Ray diffraction using Mo K$\alpha$ radiation was performed for LiMn$_{1-x}$Fe$_{x}$PO$_{4}$ with $x$ = 0, 0.1, 0.3, 0.5, and 1. In order to minimize data bias due to absorption effects and maximize resolution, data were also collected with Ag K$\alpha$ for two crystals with $x$ = 0.3 and 0.5. In order to obtain crystals small enough for the three diffractometers employed, small splinters of irregular shape cut from bigger crystals have been used. All measurements were carried out at a stabilized temperature of 100\,K. Crystallographic details are given in the Supporting Information. Excellent data quality to high resolution was obtained in all cases. The results show that all samples are of good crystallinity and feature the same structure. For the mixed Fe/Mn crystals there is no evidence of superstructure reflections indicative of long range ordering of the Fe and Mn ions. The good agreement between the structure refinements of the data sets from different crystals measured with Mo K$\alpha$ and Ag K$\alpha$ radiation on different diffractometers indicates the high accuracy of our results and gives a good approximation of the experimental errors of this technique.

The obtained relative changes of the lattice constants due to Mn/Fe substitution are in agreement with the results of the powder XRD measurements (see Fig.~\ref{latticeConst}). Direct comparison of the data is not possible since the powder data were collected at room temperature. This however allows for estimating the thermal expansion coefficient $\alpha = \Delta L / \Delta T \cdot 1 / L$. Within the error bars, thermal expansion seems to be isotropic and doping independent. The averaged value for all compositions is $\alpha = 10(2) \cdot 10^{-6}$\,K$^{-1}$.

%%%%%%%%%%%%%%%%%%%%%%%%%%%%%%%%%%%%%%%%%%%%%%%%%%%%%%%%%%%%

\begin{figure}[!h]
\centering
\includegraphics [width=0.7\columnwidth,clip] {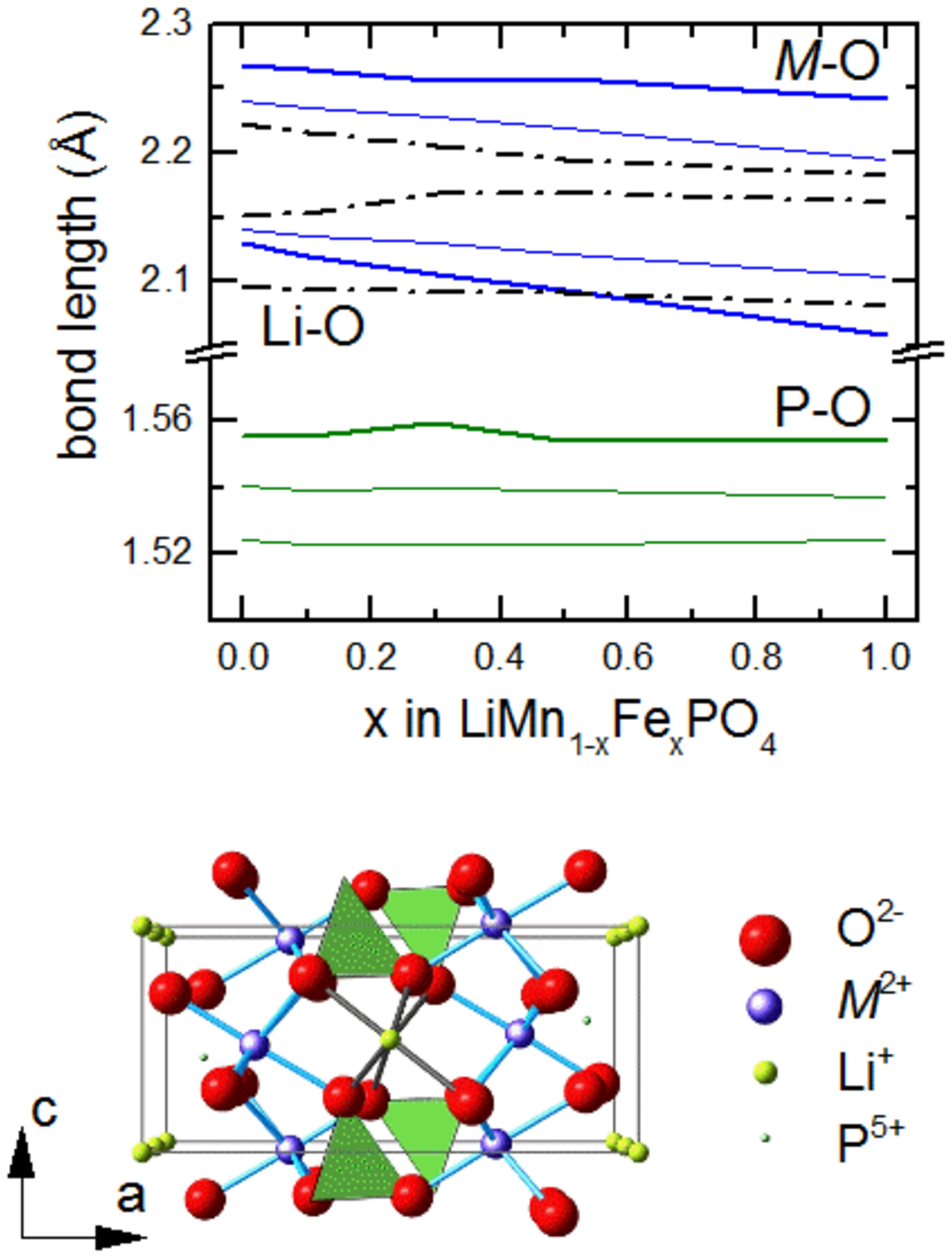}
\caption{Oxygen bond lengths and crystal structure of LiMn$_{0.5}$Fe$_{0.5}$PO$_{4}$. Bond-length multiplicity is shown by bold (2) and fine (1) lines (\textit{M}-O blue, Li-O dashed black, P-O green). PO$_4$ tetrahedra in the structure graphic are shown in green.} \label{bonds}
\end{figure}

Atomic positions and further data are given in Tab.~\ref{XRDAtomPos}. The influence of the transition metal ion $M^{2+}$, present in the respective structure, can be revealed best by considering the cation-oxygen bonds. Fig.~\ref{bonds} shows the crystal structure of LiMn$_{0.5}$Fe$_{0.5}$PO$_{4}$ along with the composition dependent oxygen-cation bond-lengths. The \textit{M}-O octahedra are rather distorted with a length difference of up to 9\,\% between the different bonds. While the mean \textit{M}-O bond length changes linearly in accordance with the ionic radii, a slight discontinuity can be observed for Li-O for an Fe content between 10\,\% and 30\,\%.
As is expected because of their strong bonds, the PO$_4$-tetrahedra are rather independent of the Mn:Fe ratio.

%%%%%%%%
\begin{table}[!h]
\centering
			 \caption{Atomic parameters (site multiplicity, Wyckoff notation, point symmetry, fractional atomic coordinates ($\cdot 10^{4}$) and equivalent isotropic displacement parameters (\AA$2^ \cdot 10^3$)). U$_{eq}$ is defined as one third of the trace of the orthogonalized U$_{ij}$ tensor. The ratio sof(Fe):sof(Mn) was fixed at the nominal value.}
			 \small
			\begin{tabular}{llllll}
			\hline
			atom & pos. & x & y & z & U$_{eq}$ \\
      \hline
      \multicolumn{6}{c}{LiMnPO$_4$, Mo K$\alpha$} \\
      \hline
      Li & 4,b,$\overline{1} $ & 0 & 0 & 5000 & 9(1) \\
      Mn & 4,c,$m$ & -2191(1) & 2500 & -285(1) & 4(1) \\
      P & 4,c,$m$ & 918(1) & 2500 & 909(1) & 4(1) \\
      O1 & 8,d,$1$ & 1612(1) & 489(1) & 2232(2) & 5(1) \\
      O2 & 4,c,$m$ & -453(1) & 2500 & 2122(2) & 5(1) \\
      O3 & 4,c,$m$ & -4041(1) & 2500 & -2690(2) & 6(1) \\
      \hline
      \multicolumn{6}{c}{LiMn$_{0.9}$Fe$_{0.1}$PO$_4$, Mo K$\alpha$} \\
      \hline
      Li & 4,b,$\overline{1} $ & 0 & 0 & 5000 & 10(1) \\
      Mn/Fe & 4,c,$m$ & -2190(1) & 2500 & -281(1) & 4(1) \\
      P & 4,c,$m$ & 920(1) & 2500 & 900(1) & 3(1) \\
      O1 & 8,d,$1$ & 1616(1) & 485(1) & 2226(1) & 5(1) \\
      O2 & 4,c,$m$ & -451(1) & 2500 & 2118(1) & 5(1) \\
      O3 & 4,c,$m$ & -4041(1) & 2500 & -2679(1) & 5(1) \\
      \hline
      \multicolumn{6}{c}{LiMn$_{0.7}$Fe$_{0.3}$PO$_4$, Mo K$\alpha$} \\
      \hline
      Li & 4,b,$\overline{1} $ & 0 & 0 & 5000 & 5(1) \\
      Mn/Fe & 4,c,$m$ & -2188(1) & 2500 & -271(1) & 4(1) \\
      P & 4,c,$m$ & 927(1) & 2500 & 879(1) & 4(1) \\
      O1 & 8,d,$1$ & 1625(1) & 480(1) & 2207(1) & 5(1) \\
      O2 & 4,c,$m$ & -446(1) & 2500 & 2102(1) & 5(1) \\
      O3 & 4,c,$m$ & -4039(1) & 2500 & -2653(1) & 5(1) \\
      \hline
      \multicolumn{6}{c}{LiMn$_{0.7}$Fe$_{0.3}$PO$_4$, Ag K$\alpha$} \\
      \hline
      Li & 4,b,$\overline{1} $ & 0 & 0 & 5000 & 7(1) \\
      Mn/Fe & 4,c,$m$ & -2188(1) & 2500 & -273(1) & 4(1) \\
      P & 4,c,$m$ & 926(1) & 2500 & 881(1) & 3(1) \\
      O1 & 8,d,$1$ & 1625(1) & 480(1) & 2209(1) & 5(1) \\
      O2 & 4,c,$m$ & -446(1) & 2500 & 2105(1) & 5(1) \\
      O3 & 4,c,$m$ & -4040(1) & 2500 & -2654(1) & 5(1) \\
      \hline
      \multicolumn{6}{c}{LiMn$_{0.5}$Fe$_{0.5}$PO$_4$, Mo K$\alpha$} \\
      \hline
      Li & 4,b,$\overline{1} $ & 0 & 0 & 5000 & 3(1) \\
      Mn/Fe & 4,c,$m$ & -2186(1) & 2500 & -264(1) & 4(1) \\
      P & 4,c,$m$ & 933(1) & 2500 & 862(1) & 4(1) \\
      O1 & 8,d,$1$ & 1633(1) & 476(1) & 2191(1) & 6(1) \\
      O2 & 4,c,$m$ & -442(1) & 2500 & 2087(1) & 5(1) \\
      O3 & 4,c,$m$ & -4037(1) & 2500 & -2630(1) & 6(1) \\
      \hline
      \multicolumn{6}{c}{LiMn$_{0.5}$Fe$_{0.5}$PO$_4$, Ag K$\alpha$} \\
      \hline
      Li & 4,b,$\overline{1} $ & 0 & 0 & 5000 & 4(1) \\
      Mn/Fe & 4,c,$m$ & -2186(1) & 2500 & -265(1) & 3(1) \\
      P & 4,c,$m$ & 933(1) & 2500 & 862(1) & 3(1) \\
      O1 & 8,d,$1$ & 1634(1) & 476(1) & 2192(1) & 5(1) \\
      O2 & 4,c,$m$ & -443(1) & 2500 & 2089(1) & 5(1) \\
      O3 & 4,c,$m$ & -4037(1) & 2500 & -2629(1) & 5(1) \\
      \hline
      \multicolumn{6}{c}{LiFePO$_4$, Mo K$\alpha$} \\
      \hline
      Li & 4,b,$\overline{1} $ & 0 & 0 & 5000 & 9(1) \\
      Fe & 4,c,$m$ & -2182(1) & 2500 & -255(1) & 10(1) \\
      P & 4,c,$m$ & 946(1) & 2500 & 818(1) & 10(1) \\
      O1 & 8,d,$1$ & 1654(1) & 464(2) & 2152(2) & 11(1) \\
      O2 & 4,c,$m$ & -435(1) & 2500 & 2060(3) & 11(1) \\
      O3 & 4,c,$m$ & -4037(1) & 2500 & -2562(3) & 11(1) \\
      \end{tabular}
	 \label{XRDAtomPos}
\end{table}
%%%%%%%%%%%%%%%%%%%%%%%%%%%%%%%%%%%%%%%%%%%%%%%%%%%%%%%%%%%%%%%%%%%%%%%%%%%%%%%%%%%%%%%%%%%%%%
\normalsize
%%%%%%%%%%%%%%%%%%%%%%%%%%%%%%%%%%%%%%FIG%%%%%%%%%%%%%%%%%%%%%%%%%%%%%%%%%%%%%%%%%%%%%%%%%%%%%
%%%%%%%%%%%%%%%%%%%%%%%%%%%%%%%%%%%%%%%%%%%%%%%%%%%%%%%%%%%%%%%%%%%%%%%%%%%%%%%%%%%%%%%%%%%%%%

%%%%%%%%%%%%%%%%%%%%%%%%%%%%%%%%%%%%%%%%%%%%%%%%%%%%%%%%%%%

The single crystal structure of \lfp\ has been determined previously for hydrothermally grown crystals~\cite{Yakubovich1977} as well as in a thorough X-ray and neutron study of several flux-grown crystals~\cite{Janssen2013}. With the X-ray scattering power of Mn and Fe only differing by one electron out of 25 and 26, respectively, refinement of the manganese and iron populations presents a great challenge. For minerals, it is well known that the choice of scattering factors has a decisive influence on the refinement of the populations of different metal ions in the same crystallographic site, and, based on extensive methodological tests with rock-forming minerals, ionic scattering factors were strongly advocated~\cite{Hawthorne1995}. While trying to determine small quantities of defects within the olivine structure, Janssen \etal\ also noted the influence of the atomic scattering factors. In order to correct for the inadequacy of the conventional (neutral) scattering factors used for all atoms, these authors allowed the population of the oxygen atoms to refine to unphysical values $> 1.0$ in Ref.~\cite{Janssen2013}.

When refined with scattering factors for the neutral atoms, our data result in oxygen populations slightly larger than unity (Li\textit{M}PO$_{4.0+x}$, $x$~=~0.10 ... 0.16), too. However, application of ionic scattering factors for all atoms except P (i.e. Li$^{+}$, Mn$^{2+}$, Fe$^{2+}$ and O$^{2-}$) give essentially stoichiometric oxygen populations, corresponding to LiFePO$_{3.99(1)}$ and LiMnPO$_{3.985(8)}$, respectively. Hence, with these scattering factors the unphysical workaround of Janssen \etal\ is avoided and in all calculations we therefore kept the site occupation factors of the oxygen atoms fixed at their stoichiometric values (corresponding to Li\textit{M}PO$_{4.00}$).

%%%%%%%%
\begin{table*}[!h]
\centering
			 \caption{Stoichiometry and Li$^+$/\textit{M}$^{2+}$ disorder as determined by single crystal XRD refinement. Fe:Mn ratio by EDX measurements. \textit{a}: Mo K$\alpha$ radiation, \textit{b}: Ag K$\alpha$ radiation, \textit{c}: Only Li$^+$/Fe$^{2+}$ exchange was considered.}
			\begin{tabular}{lllll}
			\hline
			assumed form. LiMn$_{1-x}$Fe$_x$PO$_4$ & x (nom.)& x (EDX) & x (SCD) & antisite disorder \\
			\hline
      LiMnPO$_4$$^a$ & 0 &  0.002(2) & 0.11(5) & (Li$_{0.977(4)}$Mn$_{0.011(2)}$)MnPO$_4$ \\
      LiMn$_{0.9}$Fe$_{0.1}$PO$_4$$^a$ & 0.1 & 0.108(4) & 0.09(2) & - \\
      LiMn$_{0.8}$Fe$_{0.2}$PO$_4$$^a$ & 0.2 & 0.195(4) & - & - \\
      LiMn$_{0.7}$Fe$_{0.3}$PO$_4$$^a$ & 0.3 & 0.302(5) & 0.27(2) & - \\
      LiMn$_{0.7}$Fe$_{0.3}$PO$_4$$^b$ & 0.3 & 0.302(5) & 0.36(2) & - \\
      LiMn$_{0.5}$Fe$_{0.5}$PO$_4$$^a$ & 0.5 & 0.525(5) & 0.53(2) & (Li$_{0.952(2)}$Fe$_{0.024(1)}$)(Fe$_{0.5}$Mn$_{0.5}$)PO$_4$ $^c$ \\
      LiMn$_{0.5}$Fe$_{0.5}$PO$_4$$^b$ & 0.5 & 0.525(5) & 0.45(2) & - \\
      LiFePO$_4$$^a$ & 1 & 1.00(1) & 0.95(5) & (Li$_{0.953(5)}$Fe$_{0.023(2)}$)FePO$_4$ \\
 		\end{tabular}
	 \label{tabstoich}
\end{table*}
%%%%%%%%%%%%%%%%%%%%%%%%%%%%%%%%%%%%%%%%%%%%%%%%%%%%%%%%%%%%%%%%%%%%%%%%%%%%%%%%%%%%%%%%%%%%%%

Refinement of the Fe:Mn populations was attempted, with the sum of the populations constrained to unity. The results are summarized in Tab.~\ref{tabstoich}. For the crystals containing 10 to 50\,\% Fe, the refined populations agree quite well with the Fe:Mn ratio as determined by EDX analysis (see Tab.~\ref{tabstoich}), regardless of the wavelength used to collect the diffraction data. Even in the crystals with only Fe or Mn present, the fraction of the actually absent metal (Mn and Fe, respectively) refines to zero within about two standard deviations.\footnote{We note that with LiMnPO$_4$ and LiFePO$_4$ data were only collected to less resolution (0.67\,\AA ) as with the mixed metal compounds (0.5~...~0.4\,\AA ). As high angle (high resolution) data are dominated by the heaviest atoms, the reliability of a metal population refinement is expected to increase with increasing resolution of the dataset.}

In order to estimate the presence of cation antisite disorder, i.e. the exchange of Li and Fe/Mn ions in their respective positions~\cite{Gardiner2010}, disorder according to the equation 2[Li$^{+}$]$^b \rightarrow$ [\textit{M}$^{2+}$]$^b$ was studied for the end members of the series (\textit{M}~=~Mn and \textit{M}~=~Fe) and for LiFe$_{0.5}$Mn$_{0.5}$PO$_4$.~\footnote{b and c refer to the Li$^+$ and \textit{M}$^{2+}$ positions (Wyckoff notation for $Pnma$) in the ideal crystal.} To obtain a stable refinement, the population of [\textit{M}$^{2+}$]$^c$ was fixed at the ideal value of 1.00 (\textit{sof} 0.50). This effectively results in a change of stoichiometry, as some Li$^+$ is replaced by additional \textit{M}$^{2+}$, but  appears justified if only a small fraction of ions actually change place. The results, also presented in Tab.~\ref{tabstoich}, indicate a small amount of such disorder, which may be more pronounced for \textit{M}~=~Fe than for \textit{M}~=~Mn.

\clearpage
\section{Magnetic properties}

\begin{figure*}[!h]
\centering
\includegraphics [width=0.95\columnwidth,clip] {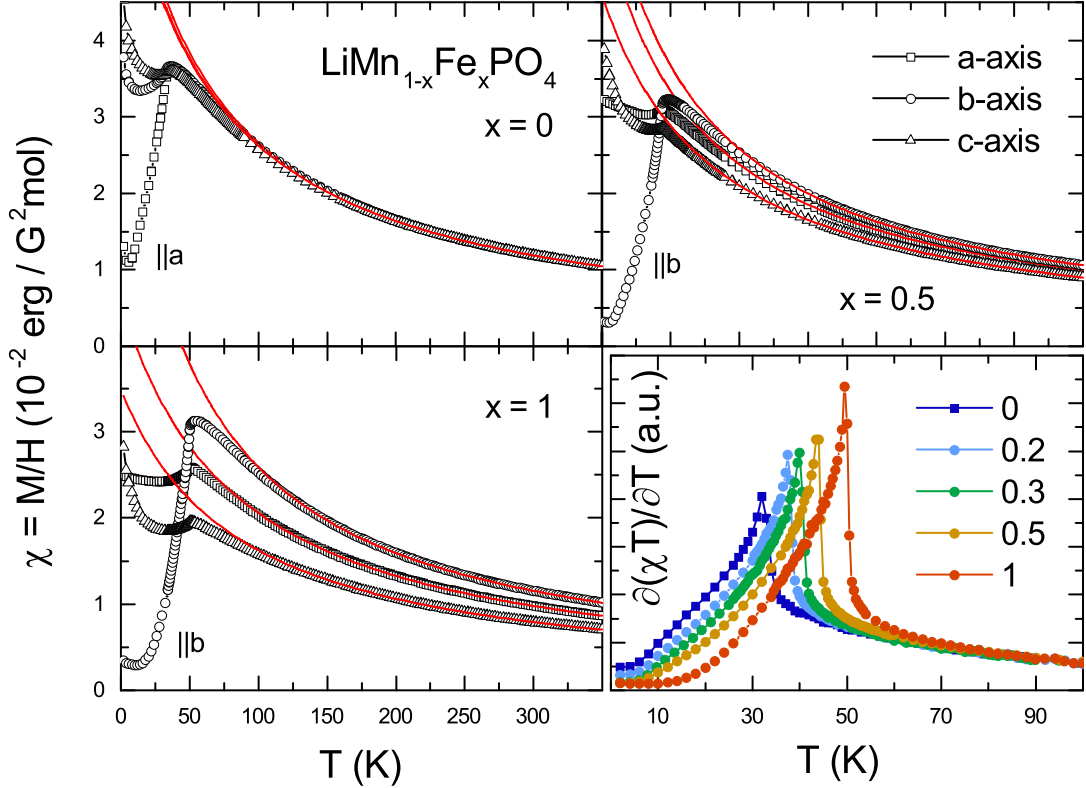}
\caption{Magnetic susceptibility $\chi = M/H$ of \lmfpx\ ($x$~=~0, 0.5, 1) measured along the main crystallographic directions (symbols). The lines show Curie-Weiss-like fits to the data. Bottom right: Magnetic specific heat $\partial (\chi_{\parallel} T)/ \partial T$ for the doping levels $x=0,0.2,0.3,0.5,$ and $1$ obtained from measurements along the magnetic easy axes.} \label{suszept}
\end{figure*}

The cuboidal crystals were used to study the magnetic properties of \lmfpx\ by means of SQUID magnetometry. The temperature dependence of the static magnetic susceptibility $\chi = M/H$ measured parallel to the main crystallographic axes is shown in Fig.~\ref{suszept} for the examples of $x$~=~0, 0.5, 1 (for x~=~0.2, 0.3 see the supporting information).

%%%%%%%%%%%%%%%%%%%%%%%%%%%%%%%%%%%%%%FIG%%%%%%%%%%%%%%%%%%%%%%%%%%%%%%%%%%%%%%%%%%%%%%%%%%%%%
%%%%%%%%%%%%%%%%%%%%%%%%%%%%%%%%%%%%%%%%%%%%%%%%%%%%%%%%%%%%%%%%%%%%%%%%%%%%%%%%%%%%%%%%%%%%%%

The susceptibilities of all crystals and along all directions exhibit maxima at low temperatures which indicate antiferromagnetic interactions between the 3$d$-transition metal moments. A steep decrease is observed below the maximum for one crystallographic direction, respectively, identifying the magnetic easy axis. In \lmp , the spins are aligned along the crystallographic $a$-axis in the long-range antiferromagnetically ordered phase. Above \tn , i.e. in the paramagnetic phase, no anisotropy is found as it is expected due to the spin-only character and associated negligible single-ion anisotropy of the magnetic Mn$^{2+}$-moments~\cite{Li2009}. The results are in agreement with single crystal data in Ref.~\cite{Wizent2009,Rudisch2013}.~\footnote{A different easy axis reported in Ref.~\cite{Wizent2009} is due to inconsistent labelling therein.}

Fe-doping yields a change in the effective magnetic anisotropy as for $x \geq 0.2$ the $b$-axis becomes the magnetic easy axis. Concomitantly, there is growing anisotropy in the paramagnetic regime, i.e. the paramagnetic $g$-factor becomes anisotropic, too. In general, this behaviour is explained in terms of large spin-orbit coupling and single-ion anisotropy in Fe$^{2+}$, which in \lfp\ were reported to be of the same order of magnitude as compared to the exchange interaction between neighboring magnetic moments~\cite{Li2006,Liang2008}.

\begin{figure}[!h]
\centering
\includegraphics [width=0.7\columnwidth,clip] {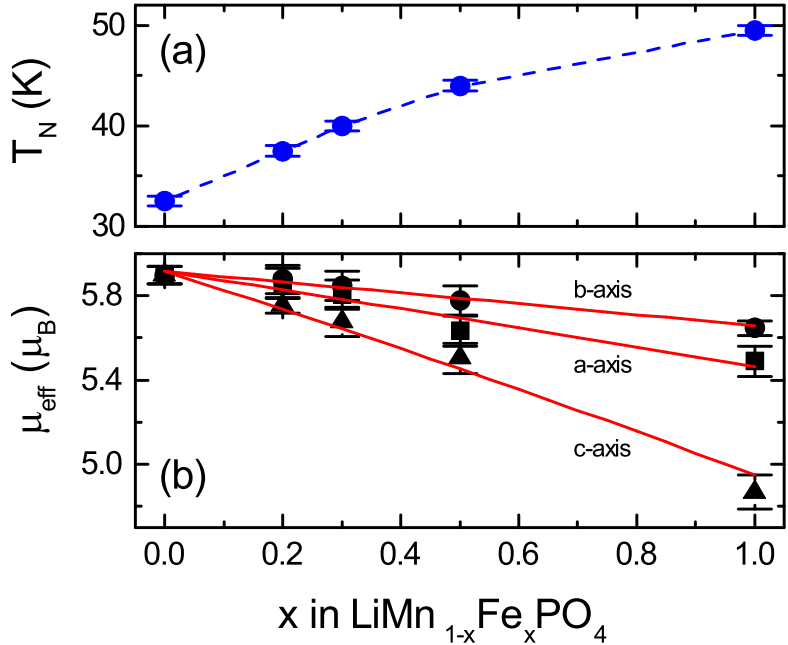}
\caption{Doping dependence of (a) \tn\ and (b) the effective magnetic moment $\mu _{\rm eff}$. Black symbols show the data, red lines reflect the averaged $g$-factors employing Eq.~\ref{eqmue}.}\label{mueff}
\end{figure}

The onset of long-range magnetic order is clearly visible if the magnetic specific heat~\cite{Fisher1962,Nordblad1981} $\partial (\chi_{\parallel} T)/ \partial T$ is considered. For all samples under study, the magnetic specific heat exhibits sharp $\lambda$-like anomalies signalling the onset of antiferromagnetic order, i.e. $T_{\rm{N}}$ (Fig.~\ref{suszept}). Note, that the thermodynamic signature of the magnetic phase transitions is by about 3.5(5)~K below the temperature of the maximum in $\chi$. \tn\ significantly increases upon Fe-doping with the doping dependence $T_{\rm{N}}(x)$ being non-linear (Fig.~\ref{mueff} (a)). This is in contrast to very weak effects of Ni-doping on \tn\ in \lmnp ~\cite{Ottmann2015}.

At high temperatures, the magnetic susceptibility obeys a Curie-Weiss-like behaviour. Fitting the data by means of $\chi = C_m / (T - \theta ) + \chi _0$ with $C_m$ being the molar Curie-constant, $\theta$ the Weiss-temperature, and $\chi _0$ a temperature independent term describes the experimental data well. From the Curie-constant, the effective magnetic moments $\mu _{\rm eff} = 2.82 \cdot C_m ^{1/2}$ are extracted (see Tab.~\ref{tabmag}). For \lmp , the data show an isotropic magnetic moment of 5.90(4)\,$\mu _{\rm{B}}$/f.u. Considering the spin moment $S=5/2$ of Mn$^{2+}$-ions, this corresponds to a $g$-factor of $g_{\rm Mn}$~=~2.01(3) which is typical for Mn$^{2+}$-ions in octahedral environment. Upon Fe-doping, the paramagnetic $g$-factors become both larger and anisotropic, with $g_{\rm c}<g_{\rm a}<g_{\rm b}$. For the further analysis, we assumed $g_{\rm Mn}$ being independent of the doping level and considered the Fe$^{2+}$-ions in the high-spin state, i.e. $S=2$. This allows extracting the anisotropic $g$-factors of the Fe$^{2+}$-ions by means of:
\begin{equation}
%\begin{split}
\mu _{\rm eff} (x) = [(1-x) \cdot g_{\rm Mn}^2 S_{\rm Mn}(S_{\rm Mn}+1) + x \cdot g_{\rm Fe}^2 S_{\rm Fe}(S_{\rm Fe}+1)]^{1/2}\label{eqmue}
%\end{split}
\end{equation}\label{effMom}
The fit to the data obtained at different $x$ yields $g_{\rm Fe}^a = 2.23(3)$, $g_{\rm Fe}^b = 2.31(2)$, $g_{\rm Fe}^c = 2.02(6)$, which are typical values for high-spin Fe$^{2+}$-ions in a distorted sixfold oxygen environment~\cite{Krzystek2006} and exposes the contributions of spin-orbit coupling and crystal field effects. Within the error bars $\mu _{\rm eff} (x)$ of all compositions can be described by the averaged values $g_{\rm Fe}$ (see Fig.~\ref{mueff} (b)). The results for \lfp\ are in agreement with single crystal measurements in Ref.~\cite{Liang2008}.

%%%%%%%%
\begin{table}[!h]
\centering
			 \caption{Parameters of the Curie-Weiss fits and N\'eel-temperatures.}
			\begin{tabular}{lcl|l}
			\hline
			$x$ & $\mu _{eff}$ ($\mu _{\rm{B}}$) a,b,c & $\theta _{mean}$ (K) & T$_{\rm{N}}$ (K) \\
      \hline
      0 &  5.90(4) & -65(5) & 32.5(5)\\
      0.2 & 5.86(7), 5.88(7), 5.75(3) &  -69(8) & 37.5(5)\\
      0.3 &  5.81(7), 5.85(7), 5.68(7)  &  -71(5) & 40.0(5)\\
      0.5 &  5.63(7), 5.78(7), 5.50(7)  &  -74(10) & 44.0(5)\\
      1 &  5.49(7), 5.65(4), 4.87(8)  &  -78(15) & 49.5(5)\\
 		\end{tabular}
	 \label{tabmag}
\end{table}
%%%%%%%%%%%%%%%%%%%%%%%%%%%%%%%%%%%%%%%%%%%%%%%%%%%%%%%%%%%%%%%%%%%%%%%%%%%%%%%%%%%%%%%%%%%%%%

%%%%%%%%%%%%%%%%%%%%%%%%%%%%%%%%%%%%%%FIG%%%%%%%%%%%%%%%%%%%%%%%%%%%%%%%%%%%%%%%%%%%%%%%%%%%%%
%%%%%%%%%%%%%%%%%%%%%%%%%%%%%%%%%%%%%%%%%%%%%%%%%%%%%%%%%%%%%%%%%%%%%%%%%%%%%%%%%%%%%%%%%%%%%%

The Weiss-temperatures $|\theta|$ significantly exceed $T_{\rm{N}}$ (see Tab.~\ref{tabmag}). In the frame of geometrically frustrated magnetic systems, the associated frustration parameter $f = |\theta|/T_{\rm{N}} \approx 2$ would indicate weak magnetic frustration in \lmfpx . This is agreement with recent inelastic neutron data on \lfp\ and \lmp , respectively, in Refs.~\cite{Li2006,Li2009,Toft-Petersen2015}, which have shown competing antiferromagnetic nearest and next-nearest-neighbor interactions in the $ab$-plane.

\section{Conclusions}

Mm-sized single crystals of the series \lmfp\ have been grown by the floating zone technique under a high pressure Ar atmosphere. The growth process is governed by the image furnace geometry and the light absorption capabilities of the melts. Upon parameter optimization, the crystal/melt interface, the temperature of the melt and the growth speed were adjusted for each composition. The structure and elemental composition of the crystals show an almost solid solution behaviour between \lmp\ and \lfp\ characterized by a linear change in lattice constants and well agreement between nominal and measured chemical compositions. Single crystal XRD data suggest a tendency to Li-\textit{M} antisite disorder which is more pronounced for the Fe-rich compounds.
The magnetization data show distinct anomalies at $T_{\rm{N}}$ which signal the onset of long range antiferromagnetic order. The sharp anomalies and the pronounced anisotropy of the susceptibility confirms high crystallinity of the samples. Comparing the different doping levels, the data show an increase of magnetic anisotropy and a change of the easy magnetic axis upon increasing the Fe-content in the \lmfpx\ single crystals.

\section*{Acknowledgment}
The authors thank Drs. D. Cruickshank and F. White (Rigaku Oxford Diffraction) for collecting data with the silver anode and Prof. Ch. Frampton for his hospitality at Brunel University, Uxbridge, UK. Furthermore we thank I. Glass for technical support. Financial support by the Excellence Initiative of the German Federal Government and States in the framework of the Heidelberg Graduate School for Fundamental Physics and by the Deutsche Forschungsgemeinschaft (DFG) through project KL1824/5 is gratefully acknowledged.

\end{document}